\documentclass{article}
\usepackage{spconf,amsmath,graphicx}
\usepackage{bbm}
\usepackage[hidelinks]{hyperref}
\usepackage{multirow}
\usepackage{amssymb}
\usepackage{bm}
\usepackage{pdfpages}
\usepackage{siunitx}
\usepackage{booktabs}
\usepackage[misc]{ifsym}
\usepackage[table]{xcolor}
\newcommand{\Dice}{\mathcal{D}}
\newcommand{\CE}{\mathcal{H}}
\newcommand{\Softmax}{\mathcal{S}}
\newcommand{\fEMA}{f^{\text{ema}}}
\newcommand{\Mix}{\operatorname{Mix}}

\usepackage[]{acronym}
\acrodef{DL}{Deep Learning}
\acrodef{US}{Ultrasound}
\acrodef{PCCL}{{\bf P}ixel-level and {\bf C}lass-level {\bf C}onsistency {\bf L}earning}
\acrodef{CNN}{Convolutional Neural Network}
\acrodef{ROI}{Region of Interest}
\acrodef{SSL}{Semi-Supervised Learning}
\acrodef{GT}{Ground Truth}
\acrodef{CR}{Consistency Regularization}
\acrodef{EMA}{Exponential Moving Average}
\acrodef{MAC}{Mutual Agreement Consistency}
\acrodef{KL}{Kullback–Leibler divergence}
\acrodef{MIG}{Mutual Information Gap}
\acrodef{CE}{Cross-Entropy}
\acrodef{Dice}{Dice based coefficient}

\usepackage{lipsum} % for dummy text

\title{Leveraging Information Divergence for Robust Semi-Supervised\\ Fetal Ultrasound Image Segmentation}
\name{Fangyijie Wang$^{\star \circ}$\Letter\thanks{This work was funded by Taighde \'{E}ireann – Research Ireland through the Centre for Research Training in Machine Learning (18/CRT/6183). \\ \Letter \, Corresponding author: fangyijie.wang@ucdconnect.ie.} \qquad Gu\'enol\'e Silvestre$^{\star \dagger}$ \qquad Kathleen M. Curran$^{\star \circ}$}
\address{$^{\star}$ Taighde \'{E}ireann – Research Ireland Centre for Research Training in Machine Learning, Ireland \\
$^{\dagger}$ School of Computer Science, University College Dublin, Ireland \\
$^{\circ}$ School of Medicine, University College Dublin, Ireland \\
}

\begin{document}
%\ninept
%
\maketitle
\begin{abstract}
% Maternal-fetal screening \ac{US} images have become the primary imaging modality to monitor fetal growth. \ac{DL} is considered the leading technology in image analysis in this domain. However, supervised learning \ac{DL} algorithms require a substantial collection of high-quality annotated images and sufficient computational resources for training and testing purposes, which is still a barrier to translating medical imaging solutions into low-resource settings. In this study, we present a novel semi-supervised approach for a lightweight model with $1.47~\text{M}$ parameters designed to efficiently and rapidly perform fetal \ac{US} segmentation tasks. Concretely, a lightweight convolutional neural network and a Transformer-based network are trained in a supervised manner with labelled data and in a cross-supervision manner with unlabelled data. We also enforce mutual agreement consistency between the lightweight and the Transformer-based network. Furthermore, we mixed up unlabelled data to improve the model's robustness. On average, our method achieves the best performance in two datasets by increasing the Dice score by 2.19\%, reducing the 95\% Hausdorff distance by 10.01, and decreasing the Average Surface Distance by 2.49 when only 5\% of the training data is annotated. Our method demonstrates superior performance compared to seven other semi-supervised learning methods. Our code is publicly available on GitHub.

Maternal–fetal \ac{US} is the primary modality for monitoring fetal development, yet automated segmentation remains challenging due to the scarcity of high-quality annotations. To address this limitation, we propose a semi-supervised learning framework that leverages information divergence for robust fetal \ac{US} segmentation. Our method employs a lightweight convolutional network ($1.47~\text{M}$ parameters) and a Transformer-based network, trained jointly with labelled data through standard supervision and with unlabelled data via cross-supervision. To encourage consistent and confident predictions, we introduce an information divergence loss that combines per-pixel Kullback–Leibler divergence and Mutual Information Gap, effectively reducing prediction disagreement between the two models. In addition, we apply mixup on unlabelled samples to further enhance robustness. Experiments on two fetal \ac{US} datasets demonstrate that our approach consistently outperforms seven state-of-the-art semi-supervised methods. When only 5\% of training data is labelled, our framework improves the Dice score by 2.39\%, reduces the 95\% Hausdorff distance by 14.90, and decreases the Average Surface Distance by 4.18. These results highlight the effectiveness of leveraging information divergence for annotation-efficient and robust medical image segmentation. Our code is publicly available on GitHub.

\end{abstract}
\section{Introduction}
\label{sec:intro}

\acf{US} imaging is widely used for prenatal evaluation of fetal growth, fetal anatomy, gestational age estimation, and pregnancy monitoring due to portability, low cost, and non-invasive nature \cite{Salomon:2011}. 
Accurate measurements allow for precise evaluations of fetal biometry and effective monitoring of fetal growth~\cite{Espinoza:2013,Papageorghiou:2014}. Therefore, a precise measurement of the \ac{ROI} of fetal structures, such as the head and abdomen of the fetus, is crucial for obstetricians. However, this operation is patient-specific, operator-dependent, and prone to intra-\nobreakdash and inter-user variability. This variability results in errors in fetal biometry assessments \cite{Espinoza:2013}.

Recent advancements in \acf{DL} have notably enhanced the field of fetal ultrasound image segmentation. Several studies have explored the development of lightweight DL models for \ac{US} image segmentation~\cite{Li:2025}.
Nevertheless, collecting many annotated \ac{US} images to train DL models is labor-intensive and time-consuming, requiring medical proficiency and clinical expertise for precise pixel-level labelling~\cite{Zegarra:2023}. Recent studies~\cite{Jiang:2024,Lyu:2025} have explored the applications of the \ac{SSL} framework for fetal \ac{US} image analysis. However, the scarcity of studies in this area has left it insufficiently explored. Developing a lightweight model using a semi-supervised approach presents a significant challenge, particularly when analysing fetal ultrasound images with limited labelled data.

To address this challenge, this work proposes a novel \ac{SSL} framework for fetal \ac{US} segmentation: \ac{PCCL}. Our method consists of a cross-teaching strategy between a lightweight \ac{CNN}, a Transformer network, and a Transformer teacher. The \ac{CNN} and Transformer are separately supervised by the \ac{GT} for labelled data. %We utilize predictions from unlabelled data generated by \ac{CNN}/Transformer networks to cross-supervision learn Transformer/\ac{CNN} networks, respectively. 
Moreover, we introduce a mutual agreement consistency learning strategy to ensure agreement at the pixel-class level between \ac{CNN} and Transformer. Additionally, we leverage mixup techniques on unlabelled data to allow the teacher model to apply \ac{CR} constraints, guiding the Transformer network. This framework enhances the robustness of the lightweight model. The key contributions of this paper are threefold: 
(1) We present a novel \ac{SSL} framework that integrates pseudo-label learning, mutual agreement, and interpolation consistency regularization. 
(2) This novel framework significantly enhances the utility of unlabelled data in semi-supervised fetal ultrasound segmentation.
(3) Our \ac{PCCL} framework outperforms seven existing semi-supervised methods in accurate segmentation of fetal head and abdomen ultrasound images.

\begin{figure*}[t]
\begin{center}
\includegraphics[width=.9\textwidth]{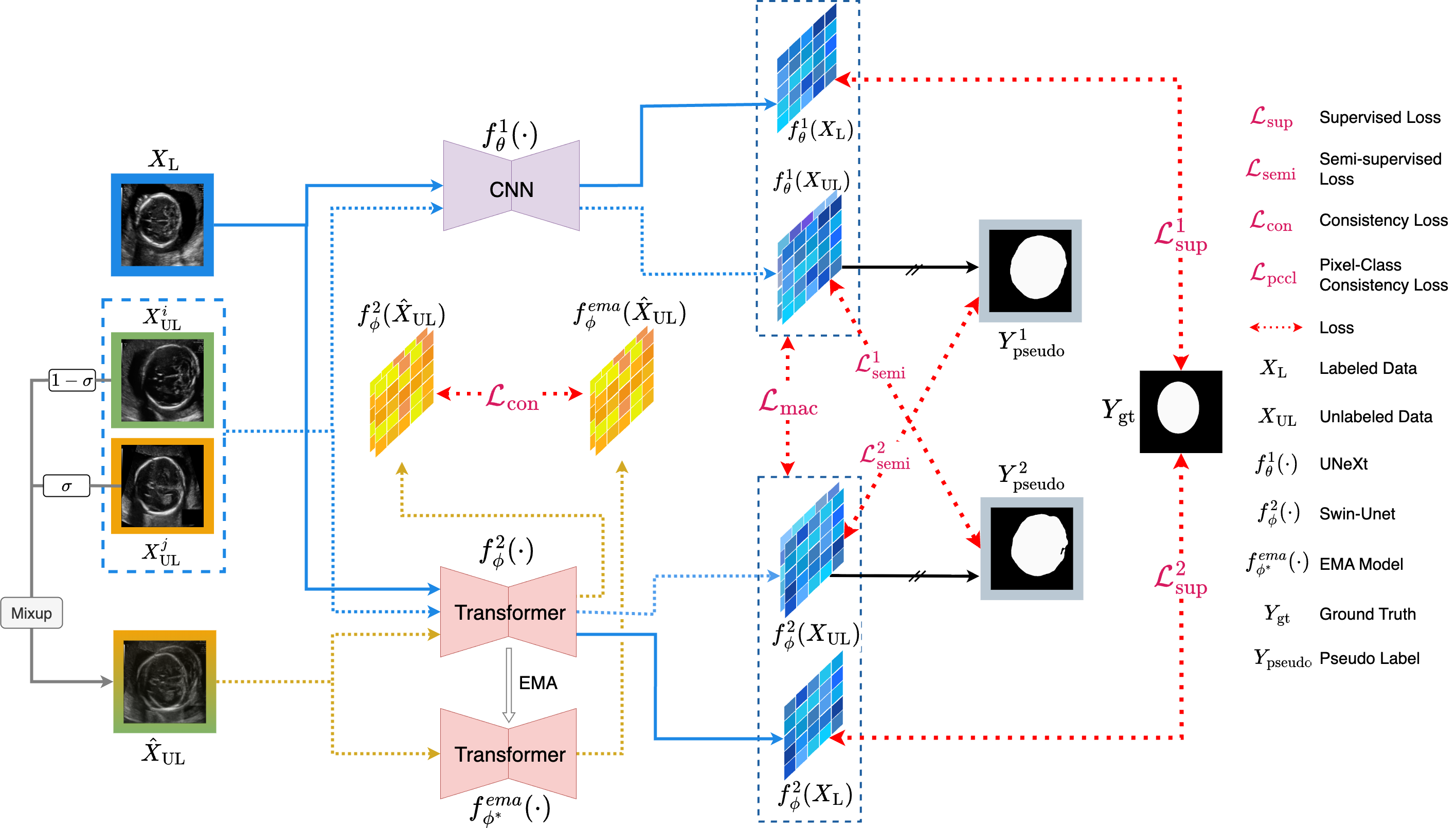}
\end{center}
    \vspace*{-1.5em}
   \caption{An overview of our \ac{PCCL} method with Swin-Unet and UNeXt for semi-supervised image segmentation.}
\label{fig:LA_SSL}
\vspace*{-1em}
\end{figure*}

\section{Method}
\label{sec:Method}

This section reviews Swin-Unet and UNeXt in Section \ref{ssec:Preliminaries}. Subsequently, we explain our semi-supervised framework \ac{PCCL} in the following sections \ref{ssec:cotrain}, \ref{ssec:ict}, \ref{ssec:pccl}, and \ref{ssec:learningloss}. The overview of our method is illustrated in Fig. \ref{fig:LA_SSL}.

% \vspace*{-1.5em}

\subsection{Preliminaries}
\label{ssec:Preliminaries}

We follow the original UNeXt~\cite{Valanarasu:2022} design to build our lightweight model $f^1_{\theta}(\cdot)$. The number of channels in the encoder path is set to $\{32,64,128,160,256\}$ to guarantee computational efficiency. 
% The hidden dimensions of the shifted multilayer perceptron (MLP) in two blocks are [$160,256$]. 
The $f^2_{\phi}(\cdot)$ is implemented using the tiny Swin-Unet architecture introduced in the original work~\cite{Cao:2023}. To fit the resolution of our input image, $448 \times 448$, the shifted window size in Swin-Unet is configured to be 7. In the encoder path of the Swin-Unet, the feature dimensions of each block are [$96, 192, 384$], while the bottleneck feature dimension is set at 768.
We initialize our model before training using pre-trained weights~\cite{Cao:2023}. Table~\ref{model_profile} compares the profiles of UNeXt and Swin-Unet. The UNeXt model shows superior lightweight and efficiency to Swin-Unet. Our \ac{PCCL} method enhances the UNeXt segmentation performance, enabling it to outperform Swin-Unet in two fetal ultrasound datasets. Further results are shown in Section~\ref{sec:ex_result}.

\begin{table}[t]
\small
\centering
\caption{\ac{CNN} and Transformer models' profiles.} %DSC: Dice Score (\%). IoU: Intersection over Union (\%). HD95: Hausdorff Distance 95\%.
\vspace*{0.2em}
\label{model_profile}
\setlength{\tabcolsep}{10pt}
\begin{tabular}{lccc}
\hline
Model & Input Size & Params $\downarrow$ & GFLOPs $\downarrow$ \\
\hline
Swin-Unet & $448\times448$ & 27.15~M & 71.17  \\
UNeXt & $448\times448$ & 1.47~M & 7.03  \\
\hline
\end{tabular}
\vspace*{-1em}
\end{table}

\subsection{Cross-supervision between \texorpdfstring{\ac{CNN}}{CNN} and Transformer}
\label{ssec:cotrain}

Inspired by existing cross-supervision works, namely Deep Co-Training \cite{Qiao:2018} and Cross Pseudo Supervision~\cite{Chen:2021}, our method \ac{PCCL} adopts a similar cross-supervision approach, combining a lightweight \ac{CNN} network with a Transformer-based network to enhance their performance mutually.
Given the unlabelled training dataset \text{UL}, the proposed framework produces two predictions: $f^1_{\theta}(\boldsymbol{X}_\text{UL})$ and $f^2_{\phi}(\boldsymbol{X}_\text{UL})$.
Based on these two predictions of $f^1_{\theta}(\cdot)$ and $f^2_{\phi}(\cdot)$, pseudo labels for the cross-pseudo-supervision strategy are generated this way:
\vspace*{-0.8em}
\[
\bm \tilde{Y}^1=f_\text{OH}\left(f^1_{\theta}\left(\boldsymbol{X}_\text{UL}\right)\right);
\tilde{Y}^2=f_\text{OH}\left(f^2_{\phi}\left(\boldsymbol{X}_\text{UL}\right)\right).
\]

No mini-batch gradient back-propagation is performed between $f^1_{\theta}(\boldsymbol{X}_\text{UL})$ and $\tilde{Y}^1$, or between $f^2_{\phi}(\boldsymbol{X}_\text{UL})$ and $\tilde{Y}^2$. $f_\text{OH}(\boldsymbol{x}\!=\!\text{c}) = \mathbbm{1}_{[\boldsymbol{x} = \text{c}]}$ denotes the one-hot encoding function used to return pseudo labels.
% \[
% f_\text{OH}(\boldsymbol{x} = \text{c}) = 
% \begin{cases}
% 1, & \text{if } \boldsymbol{x} = \text{c} \\
% 0, & \text{otherwise}
% \end{cases}
% \]
The semi-supervised loss, $\mathcal L_\text{semi}$, is the summation term $\mathcal{L}_{\text {semi}}^1 + \mathcal{L}_{\text {semi}}^2$ where
\begin{equation}
\label{semi_loss}
\begin{split}
\mathcal{L}_{\text {semi}}^1=\CE\!\left(f^1_{\theta}\left(\boldsymbol{X}_\text{UL}\right), \bm \tilde{Y}^2\right) + 
\Dice\!\left(f^1_{\theta}\left(\boldsymbol{X}_\text{UL}\right), \bm \tilde{Y}^2\right), \\
\mathcal{L}_{\text {semi}}^2=\CE\!\left(f^2_{\phi}\left(\boldsymbol{X}_\text{UL}\right), \bm \tilde{Y}^1\right) +
\Dice\!\left(f^2_{\phi}\left(\boldsymbol{X}_\text{UL}\right), \bm \tilde{Y}^1\right).
\end{split}
\end{equation}
 
\noindent $\CE(\cdot,\cdot)$ and $\Dice(\cdot,\cdot)$ denote the \ac{CE} loss and the standard \ac{Dice} loss, respectively. Within our framework, the Transformer network $f^2_{\phi}$ is used for supplementary training purposes instead of inference.

\subsection{Interpolation Consistency Learning}
\label{ssec:ict}

To further improve the usage of unlabelled data and the generalization of the Transformer model $f^2_{\phi}(\cdot)$, we encourage the predictions of an interpolation of unlabelled pixels to be consistent with the interpolation of the predictions at those pixels \cite{Verma:2022}. A mean-teacher model $\fEMA_{\phi^*}(\cdot)$ is used, where the parameters $\phi^*$ are an \ac{EMA} of the $f^2_{\phi}(\cdot)$ parameters $\phi$. During training, the parameters $\phi$ are updated to encourage consistent predictions $f^2_{\phi}(\hat{X}_\text{UL}) \!\simeq\!\Mix(\fEMA_{\phi^*}(\boldsymbol{X}^i_\text{UL}), \fEMA_{\phi^*}(\boldsymbol{X}^j_\text{UL}))$, where $\hat{\boldsymbol{X}}_\text{UL} = \Mix(\boldsymbol{X}^i_\text{UL}, \boldsymbol{X}^j_\text{UL})$. The $\Mix$ is a mixup operation with a combination ratio $\sigma = 0.5$, and $i + j$ equals the unlabelled batch size $\text{B}_\text{UL}$. The consistency loss $\mathcal{L}_\text{con}$ is defined as: 

\vspace*{-1em}
\[
\label{con_loss}
\mathcal{L}_\text{con} = \sum(f^2_{\phi}(\hat{\boldsymbol{X}}_\text{UL}) - \mathrm{Mix}(\fEMA_{\phi^*}(\boldsymbol{X}^i_\text{UL}), \fEMA_{\phi^*}(\boldsymbol{X}^j_\text{UL})))^2.
\]

\subsection{Mutual Agreement Consistency}
\label{ssec:pccl} 

In order to improve the robust consistency in semi-supervised segmentation, we introduce a new loss function called \ac{MAC} loss $\mathcal{L}_\text{mac}$. This loss function simultaneously enforces intra-pixel and class-structure agreement between the predictions of two collaborative models $f^1_{\theta}$ and $f^2_{\phi}$. The $\mathcal{L}_\text{mac}$ consists of two components: the pixel-level \ac{KL} loss \cite{Joyce:2021} and \ac{MIG} loss \cite{Chen:2018}. 
Minimizing the KL divergence reduces discrepancies in the predicted probability distributions at every spatial point, encouraging intra-pixel agreement. The \ac{MIG} loss measures the difference between inter-class similarity and intra-pixel confidence, thereby aligning the global class-wise structure within the predictions.
The proposed $\mathcal{L}_\text{mac}$ is defined as:

\vspace*{-0.75em}
\[
\label{mac_loss}
\begin{split}
\mathcal{L}_\text{mac} &= \mathrm{KL}\!\left(\Softmax\!\left(f^1_{\theta}(\boldsymbol{X}_\text{L} + \boldsymbol{X}_\text{UL})\right), \Softmax\!\left(f^2_{\phi}(\boldsymbol{X}_\text{L} + \boldsymbol{X}_\text{UL})\right)\right) \\
&+ \mathrm{MIG}\!\left(\Softmax\!\left(f^1_{\theta}(\boldsymbol{X}_\text{L} + \boldsymbol{X}_\text{UL})\right), \Softmax\!\left(f^2_{\phi}(\boldsymbol{X}_\text{L} + \boldsymbol{X}_\text{UL})\right)\right)
\end{split}
\]
where $\Softmax(\cdot)$ represents the softmax function.

% This combined loss leverages complementary constraints—local consistency at each pixel and global semantic alignment—to improve the utilization of unlabelled data in semi-supervised learning.

\subsection{The Overall Objective Function}
\label{ssec:learningloss}

All training losses are indicated by a red dashed line in Fig.~\ref{fig:LA_SSL}. The overall training objective function is a joint loss with three parts: a supervised loss, $\mathcal{L}_{\text {sup}}$, a cross-supervision loss, $\mathcal{L}_{\text {semi}}$, consistency regularization constraints from the EMA model, $\mathcal{L}_{\text {con}}$, and a pixel-class consistency loss, $\mathcal{L}_{\text {mac}}$.
The joint total loss becomes:
\begin{equation}
\label{loss}
\mathcal{L}_{\text {total}}=(\mathcal{L}_{\text {sup }}^1+\mathcal{L}_{\text {sup }}^2)+\lambda(\mathcal{L}_{\text {semi }}^1+\mathcal{L}_{\text {semi }}^2)+\tau \mathcal{L}_{\text {con}} + \beta \mathcal{L}_{\text {mac}} \nonumber
\end{equation}
where $\{\lambda$, $\tau$, $\beta\}$ are linear trade-off hyper-parameters set to $\{5.0, 1.0, 10.0\}$, respectively.
% where $\lambda$ represents the weighting factor for a Gaussian ramp-up function~\cite{Laine:2017}. This Gaussian ramp-up function facilitates the gradual transition of $f^1_{\theta}(\cdot)$ and $f^2_{\phi}(\cdot)$ from being initialized with the labelled training set to prioritizing learning from the unlabelled training set. The $\lambda$ is expressed as $\lambda = e^{-5 \times\left(1-t_{\text{iteration}} / t_{\text{maxiteration}}\right)^2}$, where $t_\text{iteration} = \text{N}_\text{epoch} / \text{B}$ and $\text{B}$ stands for the batch size.
$\mathcal{L}^1_\text{sup}$ and $\mathcal{L}^2_\text{sup}$ are the supervision losses for $f^1_{\theta}(\cdot)$ and $f^2_{\phi}(\cdot)$ based on the labelled data $\bm{X}_\text{L}$. They are designed with a combination of the \ac{Dice} and \ac{CE} losses, as follows:

\vspace*{-0.75em}
\begin{equation}
\begin{split}
\mathcal{L}_{\text {sup }}^1 &=\CE\!\left(f^1_{\theta}\left(\bm{X}_\text{L}\right), \bm{Y}_{\mathrm{gt}}\right)+\Dice\!\left(f^1_{\theta}\left(\bm{X}_\text{L}\right), \bm{Y}_{\mathrm{gt}}\right) \\
\mathcal{L}_{\text {sup }}^2&=\CE\!\left(f^2_{\phi}\left(\bm{X}_\text{L}\right), \bm{Y}_{\mathrm{gt}}\right)+\Dice\!\left(f^2_{\phi}\left(\bm{X}_\text{L}\right), \bm{Y}_{\mathrm{gt}}\right)
\end{split}
\end{equation}

\noindent Semi-supervision losses $\mathcal{L}_{\text {semi}}^1$ and $\mathcal{L}_{\text {semi}}^2$ are given in (\ref{semi_loss}). The consistency learning loss ($\mathcal{L}_{\text {con}}$) is defined in Section~\ref{con_loss} and the pixel-class consistency loss ($\mathcal{L}_{\text {mac}}$) in Section~\ref{mac_loss}.

\section{Experiments and Results}
\label{sec:ex_result}

\subsection{Datasets}
\label{ssec:data}

We used two public datasets in this study: the HC18 dataset, collected with General Electric ultrasound devices in the Netherlands \cite{Heuvel:2018_b}, and the F-Abd dataset, acquired by novice users with a low-cost portable probe connected to a smartphone in low-income countries \cite{Sappia:2025}.
The HC18 dataset comprises 999 annotated fetal head images, while the F-Abd dataset includes 187 cases with 2141 images annotated with optimal planes for abdominal circumference measurement.
Both datasets exclude cases with multiple pregnancies, congenital malformations, and aneuploidies.
Humans manually annotate the \ac{ROI} in all images. To assess \ac{SSL} segmentation performance within limited settings, we employ a 50:50 ratio to divide both datasets into training and testing sets, see Table~\ref{data_profile}. To allow pre-trained Transformer models $f^2_{\phi}$ and $\fEMA_{\phi^*}$ to be employed within our framework, we convert all \ac{US} images to RGB format.

\begin{table}
\small
\centering
\caption{The details of HC18 and F-Abd datasets.}
\vspace*{0.35em}
\label{data_profile}
\setlength{\tabcolsep}{7pt}
\begin{tabular}{lcccc}
\hline
Name & Image Size & \#Train & \#Validation & \#Test \\
\hline
HC18 & $800\times540$ & 500 & 50 & 449 \\
F-Abd & $744\times562$ & 1084 (94) & 135 (17) & 906 (76)\\
\hline
\end{tabular}
\vspace*{-1em}
\end{table}

We applied data augmentations to the labelled training data $\bm X_\text{L}$. These augmentation techniques include rotation within range $(-20^\circ, 20^\circ)$ with probability $\mathcal{P}(\cdot)=0.5$, random brightness contrast with $\mathcal{P}(\cdot)=0.5$, random blur with probability $\mathcal{P}(\cdot)=0.3$, and gaussian noise with probability $\mathcal{P}(\cdot)=0.3$. The training and testing images are resized to $448 \times 448$ to facilitate computational resource demands.

\subsection{Implementation Details}

We evaluated the performance of the proposed approach on two segmentation model architectures, Swin-Unet~\cite{Cao:2023} and UNeXt~\cite{Valanarasu:2022}. For the Swin-Unet model, we used pre-trained weights~\cite{Cao:2023} within the encoder. Our \ac{PCCL} was trained for 400 epochs, a labelled batch size of size 1 and an unlabelled batch size of size 4. A stochastic gradient descent optimizer was used with an initial learning rate of 0.01 and a momentum value of 0.9. The weight decay was set to 0.0001. Our code is developed in Python (3.11.5) using PyTorch (2.1.2) and CUDA (12.2) using one NVIDIA RTX~4090 GPU. The network was evaluated on the validation set every epoch, and the weight of the UNeXt is saved when the performance on validation outperforms the best previous performance. The above setting is also directly applied to all other baseline methods without any modification.

\begin{table}[t]
\small
\centering
\caption{The quantitative results on the HC18 dataset. The best results are in {\bf bold}. The $2^\text{nd}$ best results are in \underline{underline}. %DSC: Dice Score (\%). IoU: Intersection over Union (\%). HD95: Hausdorff Distance 95\%.
}
\label{hc18_res}
\vspace*{0.3em}
\setlength{\tabcolsep}{4pt}
\begin{tabular}{lccccc}
\hline
\multirow{2}{*}{Method} & \#L/ & \#UL/ & \multirow{2}{*}{DSC $\uparrow$} & \multirow{2}{*}{HD95 $\downarrow$} & \multirow{2}{*}{ASD $\downarrow$} \\
% \cmidrule{4-6}
& \#Case & \#Case & & & \\
\hline
\rowcolor{lightgray} \multicolumn{6}{c}{HC18} \\
\hline
UNeXt & 500 & \multirow{2}{*}{-} & 93.37 & 39.22 & 13.43 \\
Swin-Unet & (100\%) & & 96.94 & 9.80 & 3.95 \\
% \hline
UNeXt & 25 & \multirow{2}{*}{-} & 38.79 & 100.19 & 35.82 \\
Swin-Unet & (5\%) &  & 93.60 & 30.24 & 9.93 \\
\hline
%Fully-supervised & &  &  &  &  &  & \\
MT~\cite{Tarvainen:2017} & \multirow{8}{*}{25} & \multirow{8}{*}{475} & 89.43 & 55.89 & 19.23 \\
DAN~\cite{Zhang:2017} & \multirow{8}{*}{(5\%)} & \multirow{8}{*}{(95\%)} & 91.64 & 85.66 & 56.17 \\
% DCT~\cite{Qiao:2018} & & & 91.69 & 85.91 & 54.29 \\
UAMT~\cite{Yu:2019} &  &  & 90.97 & 52.10 & 17.66 \\
CPS~\cite{Chen:2021} & & & 91.35 & 45.49 & 15.75 \\
ICT~\cite{Verma:2022} & & & 91.64 & 46.70 & 15.97 \\
CTCT~\cite{Luo:2022} & & & 91.50 & 44.18 & 15.22 \\
PCPCS~\cite{Ma:2024}  & & & \underline{92.26} & \underline{41.70} & \underline{14.10} \\
\ac{PCCL}(\bf Ours) & & & \textbf{94.34} & \textbf{24.44} & \textbf{8.50} \\
\hline\hline
% UNeXt & 50 & \multirow{2}{*}{-} & 82.72 & 81.94 & 29.76 \\
% Swin-Unet & (10\%) &  & 95.36 & 17.53 & 6.42 \\
% \hline
% %Fully-supervised & &  &  &  &  &  & \\
% MT~\cite{Tarvainen:2017} & \multirow{6}{*}{50} & \multirow{6}{*}{450} & 94.38 & 30.39 & 10.34 \\
% % DAN~\cite{Zhang:2017} & & & 92.04 & 86.27 & 53.77  \\
% % DCT~\cite{Qiao:2018} & & & 94.97 & 91.06 & \underline{29.73} \\
% UAMT~\cite{Yu:2019} & \multirow{6}{*}{(10\%)} & \multirow{6}{*}{(90\%)} & 94.86 & 27.97 & 9.56 \\
% CPS~\cite{Chen:2021} & & & 94.61 & 27.75 & 9.43 \\		
% ICT~\cite{Verma:2022} & & & 94.70 & 27.81 & 9.81 \\
% CTCT~\cite{Luo:2022} & & & 94.80 & 90.75 & 35.80 \\
% PCPCS~\cite{Ma:2024}  & & & \underline{94.14} & \underline{27.20} & \underline{9.77} \\
% \ac{PCCL}(\bf Ours) & & & \textbf{94.82} & \textbf{25.76} & \textbf{8.97} \\
% \hline
\rowcolor{lightgray} \multicolumn{6}{c}{F-Abd} \\
\hline
UNeXt & 1084/94 & \multirow{2}{*}{-} & 88.07 & 45.68 & 15.43 \\
Swin-Unet & (100\%) & & 93.09 & 15.80 & 6.34 \\
% \hline
UNeXt & 29/5 & \multirow{2}{*}{-} & 56.27 & 139.05 & 58.02 \\
Swin-Unet & (5\%) &  & 64.84 & 80.12 & 34.52 \\
\hline
MT~\cite{Tarvainen:2017} & \multirow{7}{*}{29/} & \multirow{7}{*}{1055/} & 61.31 & 112.59 & 46.62 \\
DAN~\cite{Zhang:2017} & \multirow{7}{*}{5} & \multirow{7}{*}{89} & 54.20 & 131.64 & 56.76 \\
UAMT~\cite{Yu:2019} & \multirow{7}{*}{(5\%)} & \multirow{7}{*}{(95\%)} & 63.82 & 110.79 & 45.63 \\
CPS~\cite{Chen:2021} & & & \underline{64.71} & 101.26 & 43.11 \\
ICT~\cite{Verma:2022} & & & 63.17 & 127.33 & 51.58 \\
CTCT~\cite{Luo:2022} & & & 64.15 & 98.93 & 42.34 \\
PCPCS~\cite{Ma:2024}  & & & 63.99 & \underline{87.35} & \underline{38.05} \\
\ac{PCCL}(\bf Ours) & & & \textbf{66.68} & \textbf{74.82} & \textbf{35.29} \\
\hline
\end{tabular}
\vspace*{-1em}
\end{table}

\vspace*{-1em}
\subsection{Results}

We compared \ac{PCCL} with seven representative \ac{SSL} methods using only 5\% of the labelled training data. Performance was evaluated using Dice score (DSC), 95\% Hausdorff distance (HD95), and Average Surface Distance (ASD). As shown in Table \ref{hc18_res}, \ac{PCCL} improves the DSC by 2.41\%, while reducing HD95 by 7.49 and ASD by 2.21 on the HC18 dataset. Significantly, \ac{PCCL} also surpasses the fully supervised UNeXt model on all metrics, underscoring its ability to leverage unlabelled data more effectively than conventional supervised training.
For the F-Abd dataset, \ac{PCCL} achieves the best performance among all semi-supervised methods. Nevertheless, a noticeable performance gap remains between semi-supervised and fully supervised approaches. This discrepancy can be attributed to the limited number of labelled samples used in our experiments (only 5 cases) and the use of low-cost portable probes, which generate noisy images that are susceptible to both intra-\nobreakdash and inter-user variability.

\begin{figure}
\begin{center}
\includegraphics[width=.8\linewidth]{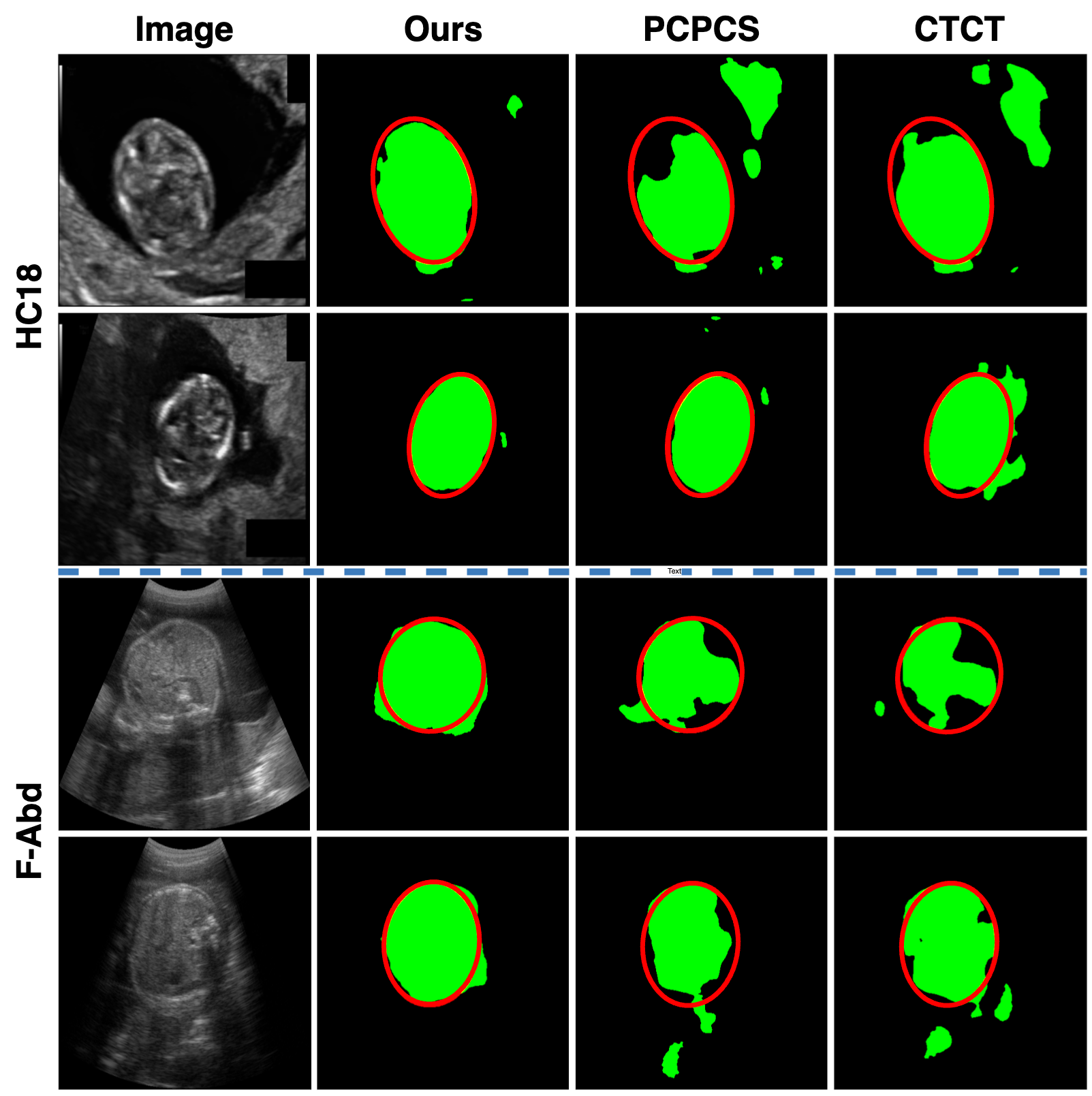}
\end{center}
\vspace*{-1em}
   \caption{Visual comparison of the top three methods. GT is in red, and the predicted results are in green.}
\label{fig:res_vis}
\end{figure}

Fig.~\ref{fig:res_vis} compares our \ac{PCCL}'s predictions with GT masks and the top 2 semi-supervised methods. It demonstrates that \ac{PCCL} yields more reliable predictions for segmenting fetal head and abdomen \ac{ROI}s across two datasets. This result demonstrates the efficacy of our \ac{PCCL} in accurately segmenting fetal organs in \ac{US} images acquired from high- and low-cost devices across diverse geographical regions.\\[-0.5em]

\noindent {\bf Ablation Study:} Table \ref{ablation} presents the results of our ablation study. Comparing the first and second rows shows that incorporating the joint loss $\mathcal{L}_{\text {semi}} + \mathcal{L}_{\text {con}}$ improves the performance of model $f^1_{\theta}$. The comparison between the first and third rows further demonstrates that the $\mathcal{L}_{\text {mac}}$ effectively enforces pixel-class consistency, leading to better segmentation. Finally, the last row shows that combining all three losses yields the best performance on fetal head ultrasound segmentation, confirming the overall effectiveness of our approach.

\begin{table}
\vspace*{-0.8em}
\small
\centering
\caption{Ablation study on the HC18 dataset with 5\% labelled data. The best results are in {\bf bold}.}\label{ablation}
\vspace*{0.3em}
\setlength{\tabcolsep}{9pt}
\begin{tabular}{lll||ccc}
\hline
$\mathcal{L}_{\text {semi}}$ & $\mathcal{L}_{\text {con}}$ & $\mathcal{L}_{\text {mac}}$ & DSC $\uparrow$ & HD95 $\downarrow$ & ASD $\downarrow$ \\
\hline\hline
\checkmark &  &  & 91.40 & 44.88 & 15.59 \\
\checkmark & \checkmark & & 92.74 & 34.38 & 12.10 \\
\checkmark &  & \checkmark & 93.82 & 27.68 & 9.48 \\
\checkmark & \checkmark & \checkmark & \textbf{94.34} & \textbf{24.44} & \textbf{8.50} \\
\hline
\end{tabular}
\vspace*{-1em}
\end{table}

\vspace*{-0.6em}

\section{Conclusion}

% We introduce \ac{PCCL}, a novel SSL approach that efficiently utilizes limited annotations to train a lightweight model for fetal ultrasound segmentation. By combining the benefits of pixel-class consistency learning and a co-training strategy to optimize the usage of unlabelled data, our proposed method contributes to enhancing robust and lightweight DL models in fetal ultrasound segmentation tasks. Furthermore, the \ac{PCCL} method maximizes the utilization of unlabelled data through interpolation consistency learning, surpassing existing semi-supervised methods. To our knowledge, it achieves SOTA performance in fetal head and abdomen segmentation in ultrasound images.

We present \ac{PCCL}, a novel \ac{SSL} approach that leverages limited annotations to train a lightweight model for fetal \ac{US} segmentation. By integrating pixel-class consistency learning and a cross-supervision strategy, \ac{PCCL} optimally exploits unlabelled data and enhances lightweight model robustness under annotation scarcity in fetal ultrasound domain. Moreover, its interpolation consistency learning strategy further maximizes the utility of unlabelled samples, enabling \ac{PCCL} to consistently yield superior segmentation accuracy than seven recent semi-supervised approaches. Our work highlights the effectiveness of leveraging information divergence for robust and annotation-efficient medical image segmentation.
\bibliographystyle{IEEEbib}
\bibliography{refs}

\end{document}